# New Julia and Mandelbrot Sets for Jungck Ishikawa Iterates

**Suman Joshi[1], Dr. Yashwant Singh Chauhan[2], Dr. Ashish Negi[3]**
*[1](Computer Science & Engg. Department, G. B. Pant Engineering College Pauri Garhwal, India)*
*[2](Computer Science & Engg. Department, G. B. Pant Engineering College Pauri Garhwal, India)*
*[3](Computer Science & Engg. Department, G. B. Pant Engineering College Pauri Garhwal, India)*

***ABSTRACT :*** *The generation of fractals and study of the dynamics of polynomials is one of the emerging and interesting field of research nowadays. We introduce in this paper the dynamics of polynomials $z^n - z + c = 0$ for $n \geq 2$ and applied Jungck Ishikawa Iteration to generate new Relative Superior Mandelbrot sets and Relative Superior Julia sets. In order to solve this function by Jungck –type iterative schemes, we write it in the form of $Sz = Tz$, where the function T, S are defined as $Tz = z^n + c$ and $Sz = z$. Only mathematical explanations are derived by applying Jungck Ishikawa Iteration for polynomials in the literature but in this paper we have generated Relative Mandelbrot sets and Relative Julia sets.*

## I. INTRODUCTION

The word "Fractal" which is taken from the Latin word "fractus" meaning "broken" was given by mathematician Benoit B Mandelbrot in 1975[1] to describe irregular and intricate natural phenomenon as lunar landscapes, mountains, branches of trees and coastlines etc. Fractals are defined as "objects that appear to be broken into number of pieces and each piece is a copy of the entire shape". The object Mandelbrot set was given by Mandelbrot in 1979 and its relative object Julia set due to their beauty and complexity of their nature have become rich area of research nowadays.

In Mandelbrot's opinion, the turning point in fractal study occurred in 1970-1980 with his research of the Fatou-julia theory of iteration. This theory had last been changed in 1918. Mandelbrot used a computer to investigate a small portion of Fatou-Julia, which he referred to as the $\mu$-map. It was later renamed the Mandelbrot set (M-set) in his honor by Adrian Douady and John Hubbard.

Fixed point theorem is one of the major tools and it has its diversified applications in the theory of fuzzy mathematics, fractals, theory of games, dynamics programming etc. For a function f having a set X as both domain and range, a fixed point of f is a point x of X for which f(x) =x [2].

## II. PRELIMINARIES

**1. Ishikawa Iteration** [3] Let X be a subset of real or complex numbers and T: X→ X for $x_0 \in X$, we have the sequences $\{x_n\}$ and $\{y_n\}$ in X in the following manner:

$$\left.\begin{array}{l} x_{n+1} = \alpha_n T y_n + (1-\alpha_n) x_n \\ y_n = \beta_n T x_n + (1-\beta_n) x_n \end{array}\right\}$$

where $0 \leq \beta_n \geq 1$ and $0 \leq \alpha_n \geq 1$ and $\alpha_n$ & $\beta_n$ both convergent to non zero number.

**2. Definition** [4] The sequences $\{x_n\}$ and $\{y_n\}$ constructed above is called Ishikawa sequences of iteration or relative superior sequences of iterate. We denote it by RSO($x_0, \alpha_n, \beta_n, t$). Notice that RSO ($x_0, \alpha_n, \beta_n, t$) with $\beta_n = 1$ is RSO($x_0, \alpha_n, t$) i.e. Mann's orbit and if we place $\alpha_n = \beta_n = 1$ then RSO ($x_0, \alpha_n, \beta_n, t$) reduces to O ($x_0, t$). We remark that Ishikawa orbit RSO($x_0, \alpha_n, \beta_n, t$) with $\beta_n = 1/2$ is Relative superior orbit. Now we define Julia set for function with respect to Ishikawa iterates. We call them as Relative Superior Julia sets.

**3. Definition** [4] The set of points SK whose orbits are bounded under Relative superior iteration of function Q (z) is called Relative Superior Julia sets. Relative Superior Julia set of Q is a boundary of Julia set RSK.

**4. Jungck Ishikawa Iteration** [5] Let $(X, \|.\|)$ be a Banach space and Y an arbitrary set. Let S, T: Y→X be two non self mappings such that $T(Y) \subseteq S(Y)$, S(Y) is a complete subspace of X and S is injective. Then for $x_o \in Y$, define the sequence $\{Sx_n\}$ and $\{Sy_n\}$ iteratively by

$$\left.\begin{array}{l} Sx_{n+1} = \alpha_n Ty_n + (1-\alpha_n)Sx_n \\ Sy_n = \beta_n Tx_n + (1-\beta_n)Sx_n \end{array}\right\}$$

where n=0, 1 ….. and $0 \leq \beta_n \geq 1$ and $0 \leq \alpha_n \geq 1$ and $\alpha_n$ & $\beta_n$ both convergent to non zero number.

### III. FIGURES AND TABLES
### 5. Fixed Points
1.1 Fixed points of quadratic polynomial

**TABLE 1:** Orbit of F (z) for ($z_o$=-0.3124999945+0.7942708667i) at $\alpha$=0.5, $\beta$=0.5, c=0.1

| No. of iterations | |Sz| | No. of iterations | |Sz| |
|---|---|---|---|
| 1 | 0.3124 | 13 | 0.1123 |
| 2 | 0.0478 | 14 | 0.1124 |
| 3 | 0.0146 | 15 | 0.1125 |
| 4 | 0.0546 | 16 | 0.1126 |
| 5 | 0.0789 | 17 | 0.1126 |
| 6 | 0.0933 | 18 | 0.1126 |
| 7 | 0.1016 | 19 | 0.1126 |
| 8 | 0.1063 | 20 | 0.1126 |
| 9 | 0.1090 | 21 | 0.1127 |
| 10 | 0.1106 | 22 | 0.1127 |
| 11 | 0.1115 | 23 | 0.1127 |
| 12 | 0.1120 | 24 | 0.1127 |

Here we observe that the value converges to a fixed point after 21 iterations.

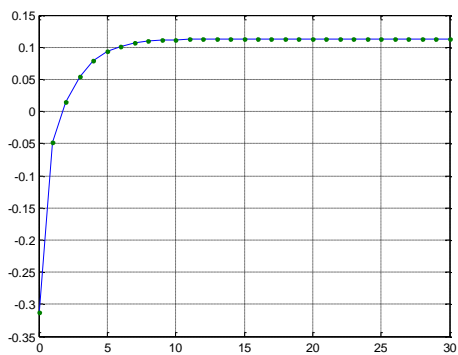

**Fig1**: Orbit of F(x) for ($z_o$=0.3124999945+0.7942708667i) at $\alpha$=0.5, $\beta$=0.5, c=0.1

**TABLE 2:** Orbit of F (z) for ($z_0$=0.275-1.625i) at $\alpha$=0.8, $\beta$=0.4, c=0.1

| No. of iterations | |Sz| | No. of iterations | |Sz| |
|---|---|---|---|
| 1 | 0.2750 | 11 | 0.1125 |
| 2 | 0.7462 | 12 | 0.1126 |
| 3 | 0.7270 | 13 | 0.1126 |
| 4 | 0.4839 | 14 | 0.1127 |
| 5 | 0.1549 | 15 | 0.1127 |
| 6 | 0.0799 | 16 | 0.1127 |
| 7 | 0.0985 | 17 | 0.1127 |
| 8 | 0.1078 | 18 | 0.1127 |
| 9 | 0.1111 | 19 | 0.1127 |
| 10 | 0.1121 | 20 | 0.1127 |

Here we observe that the value converges to a fixed point after 14 iterations.

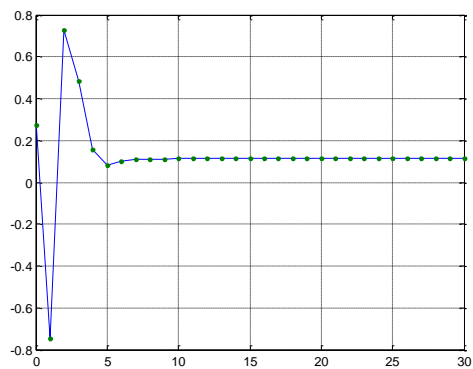

**Fig 2.** Orbit of F(x) for ($x_o$=0.275-1.625i) at $\alpha$=0.8, $\beta$=0.4, c=0.1

**TABLE 3:** Orbit of F (z) for ($z_o$=1.5-7.6i) at $\alpha$=0.1, $\beta$=0.1, c=0.1

| No. of iterations | |Sz| | No. of iterations | |Sz| |
|---|---|---|---|
| 140 | 0.1126 | 147 | 0.1127 |
| 141 | 0.1126 | 148 | 0.1127 |
| 142 | 0.1126 | 149 | 0.1127 |
| 143 | 0.1126 | 150 | 0.1127 |
| 144 | 0.1126 | 151 | 0.1127 |
| 145 | 0.1126 | 152 | 0.1127 |
| 146 | 0.1127 | 153 | 0.1127 |

We skipped 139 iterations and observed that the value converges to a fixed point after 146 iterations.

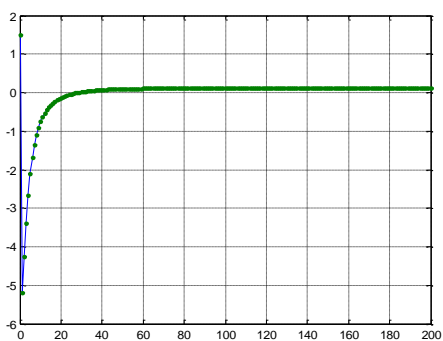

**Fig 3:** Orbit of F (z) for ($z_o$=1.5-7.6i) at $\alpha$=0.1, $\beta$=0.1, c=01

1.2 Fixed points of cubic polynomial

**TABLE 1:** Orbit of F (z) for ($z_o$=0.09375+0.2625i) at $\alpha$=0.8, $\beta$=0.8, c=0.1

| No. of iterations | |Sz| |
|---|---|
| 1 | 0.0937 |
| 2 | 0.0988 |
| 3 | 0.1005 |
| 4 | 0.1009 |
| 5 | 0.1010 |
| 6 | 0.1010 |
| 7 | 0.1010 |
| 8 | 0.1010 |
| 9 | 0.1010 |
| 10 | 0.1010 |

Here we observe that the value converges to a fixed point after 6 iterations.

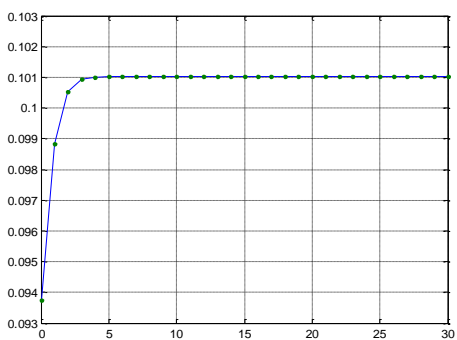

**Fig 1** Orbit of F (z) for ($z_o$=0.09375+0.2625i) at $\alpha$=0.8, $\beta$=0.8, c=0.1

**TABLE 2:** Orbit of F (z) for ($z_o$=0.025-1.3875i) at $\alpha$=0.5, $\beta$=0.5, c=0.1

| No. of iterations | |Sz| | No. of iterations | |Sz| |
|---|---|---|---|
| 1 | 0.0250 | 11 | 0.1008 |
| 2 | 0.0684 | 12 | 0.1009 |
| 3 | 0.0838 | 13 | 0.1009 |
| 4 | 0.0888 | 14 | 0.1010 |
| 5 | 0.0934 | 15 | 0.1010 |
| 6 | 0.0967 | 16 | 0.1010 |
| 7 | 0.0987 | 17 | 0.1010 |
| 8 | 0.0998 | 18 | 0.1010 |
| 9 | 0.1004 | 19 | 0.1010 |
| 10 | 0.1007 | 20 | 0.1010 |

Here we observe that the value converges to a fixed point after 16 iterations.

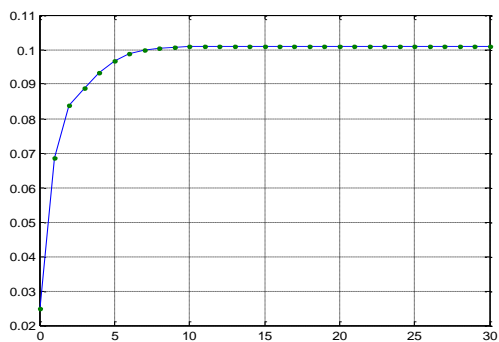

**Fig 2:** Orbit of F (z) for ($z_o$=0.025-1.3875i) at $\alpha$=0.5, $\beta$=0.5, c=0.1

**TABLE 3:** Orbit of F (z) for ($z_o$=-1.10625-0.39375i) at $\alpha$=0.4, $\beta$=0.6, c=0.1

| No. of iterations | |Sz| | No. of iterations | |Sz| |
|---|---|---|---|
| 1 | 1.10625 | 11 | 0.10116 |
| 2 | 0.13639 | 12 | 0.10111 |
| 3 | 0.12193 | 13 | 0.10108 |
| 4 | 0.11184 | 14 | 0.10106 |
| 5 | 0.10620 | 15 | 0.10105 |
| 6 | 0.10355 | 16 | 0.10104 |
| 7 | 0.10232 | 17 | 0.10104 |
| 8 | 0.10172 | 18 | 0.10103 |
| 9 | 0.10142 | 19 | 0.10103 |
| 10 | 0.10125 | 20 | 0.10103 |

Here we observe that the value converges to a fixed point after 18 iterations.

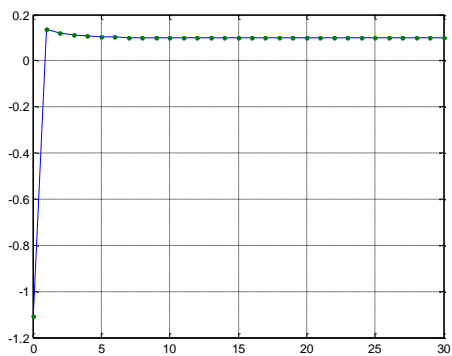

**Fig 3:** Orbit of F (z) for (z$_o$=-1.10625-0.39375i) at $\alpha$ =0.4, $\beta$ =0.6, c=0.1

1.3 Fixed points of biquadratic polynomial

**TABLE 1:** Orbit of F (z) for (z$_o$= -0.00625+0.6625i) at $\alpha$ =0.8, $\beta$ =0.8, c=0.1

| No. of iterations | |Sz| |
|---|---|
| 1 | 0.0062 |
| 2 | 0.0764 |
| 3 | 0.0953 |
| 4 | 0.0991 |
| 5 | 0.0999 |
| 6 | 0.1000 |
| 7 | 0.1000 |
| 8 | 0.1001 |
| 9 | 0.1001 |
| 10 | 0.1001 |

Here we observe that the value converges to a fixed point after 8 iterations.

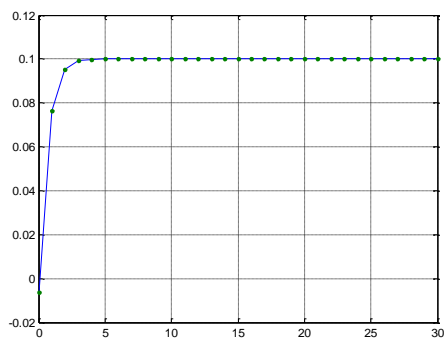

**Fig 1:** Orbit of F (z) for (z$_o$= -0.00625+0.6625i) at $\alpha$ =0.8, $\beta$ =0.8, c=0.1

**TABLE 2:** Orbit of F (z) for (z$_o$= -0.06875+1.0875i) at $\alpha$ =0.5, $\beta$ =0.5, c=0.1

| No. of iterations | |Sz| | No. of iterations | |Sz| |
|---|---|---|---|
| 1 | 0.0687 | 11 | 0.0988 |
| 2 | 0.4891 | 12 | 0.0994 |
| 3 | 0.2088 | 13 | 0.0997 |
| 4 | 0.0544 | 14 | 0.0999 |
| 5 | 0.0227 | 15 | 0.1000 |
| 6 | 0.0613 | 16 | 0.1000 |
| 7 | 0.0807 | 17 | 0.1000 |
| 8 | 0.0903 | 18 | 0.1000 |
| 9 | 0.0952 | 19 | 0.1001 |
| 10 | 0.0976 | 20 | 0.1001 |

We observe that the value converges to a fixed point after 19 iterations.

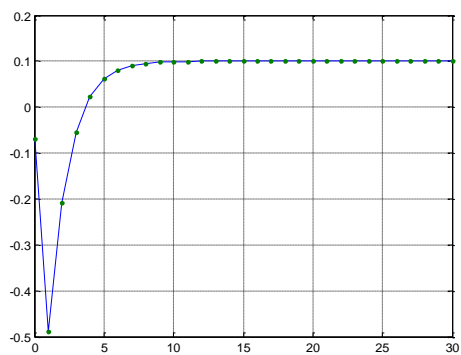

**Fig 2:** Orbit of F (z) for (z$_o$= -0.06875+1.0875i) at $\alpha$ =0.5, $\beta$ =0.5, c=0.1

**TABLE 3:** Orbit of F (z) for (z$_o$=-0.26875+1.04375i) at $\alpha$ =0.4, $\beta$ =0.6, c=0.1

| No. of iterations | \|Sz\| | No. of iterations | \|Sz\| |
|---|---|---|---|
| 1 | 0.26875 | 13 | 0.09991 |
| 2 | 0.04699 | 14 | 0.09998 |
| 3 | 0.06819 | 15 | 0.10003 |
| 4 | 0.08093 | 16 | 0.10006 |
| 5 | 0.08858 | 17 | 0.10007 |
| 6 | 0.09318 | 18 | 0.10009 |
| 7 | 0.09594 | 19 | 0.10009 |
| 8 | 0.09760 | 20 | 0.10009 |
| 9 | 0.09860 | 21 | 0.10010 |
| 10 | 0.09920 | 22 | 0.10010 |
| 11 | 0.09956 | 23 | 0.10010 |
| 12 | 0.09978 | 24 | 0.10010 |

Here we observe that the value converges to a fixed point after 21 iterations.

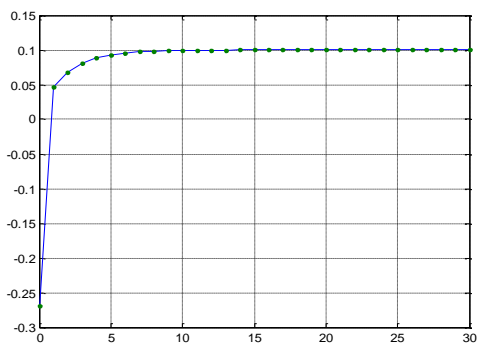

**Fig 3:** Orbit of F (z) for (z$_o$=-0.26875+1.04375i) at $\alpha$ =0.4, $\beta$ =0.6, c=0.1

### 6. Generation of Relative Superior Julia Sets

We generated the Relative Superior Julia sets. We present here some beautiful filled Relative Superior Julia sets for quadratic, cubic and biquadratic function.

1.4 Relative Superior Julia sets for Quadratic function

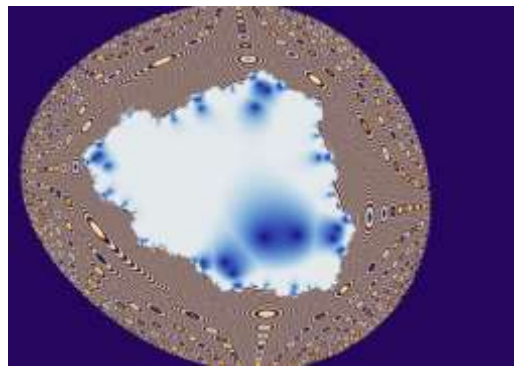

Fig 1: Relative Superior Julia Set for $\alpha = \beta$ =0.5 & c =0.1

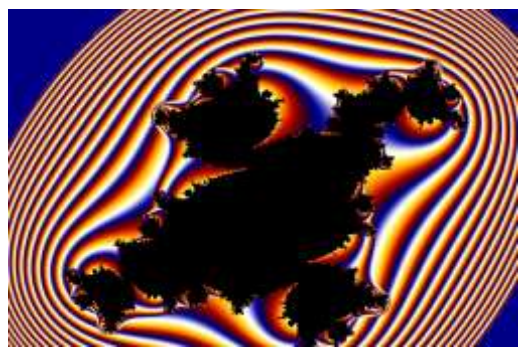

Fig 2: Relative Superior Julia Set for $\alpha$ =0.8, $\beta$ =0.4, c=0.1

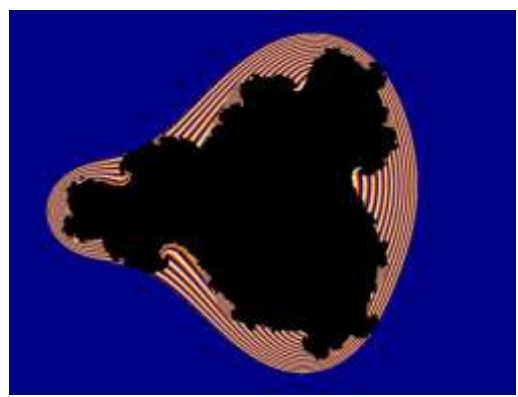

Fig 3: Relative Superior Julia Set for $\alpha$ =0.1, $\beta$ =0.1, c=0.1

1.5 Relative Superior Julia Sets for Cubic function

1.6 Relative Superior Julia sets for biquadratic function

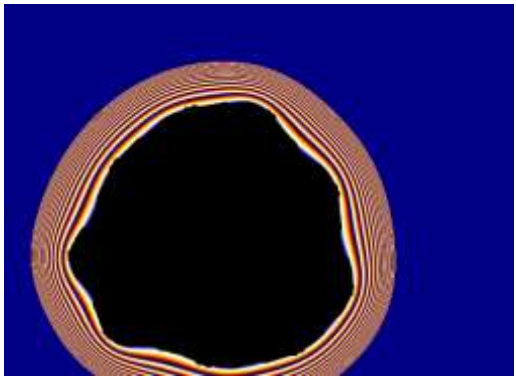

Fig1: Relative Superior Julia Set for $\alpha = \beta =0.8$, c=0.1

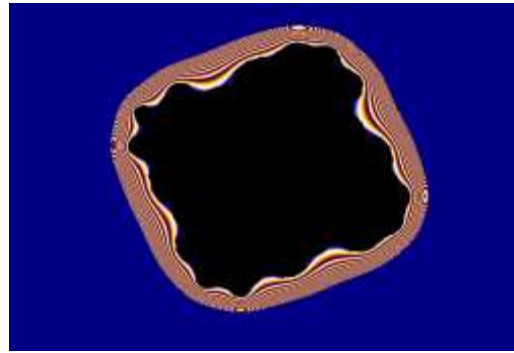

Fig 1: Relative Superior Julia Set for $\alpha = 0.8$, $\beta = 0.8$, c=0.1

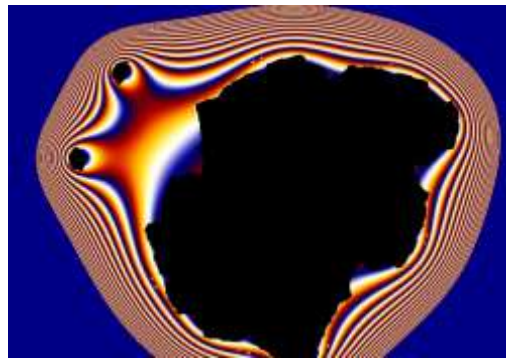

Fig 2: Relative Superior Julia Set for $\alpha =0.5$, $\beta =0.5$, c = 0.1

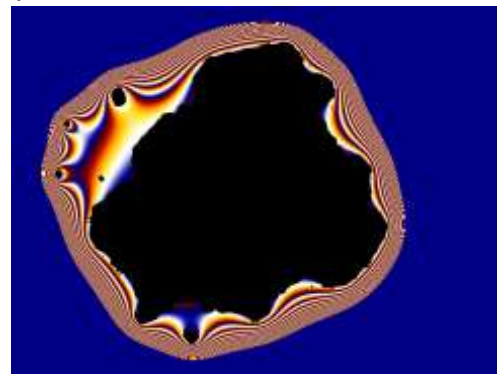

Fig 2: Relative Superior Julia Set for $\alpha = 0.5$, $\beta = 0.5$, c=0.1

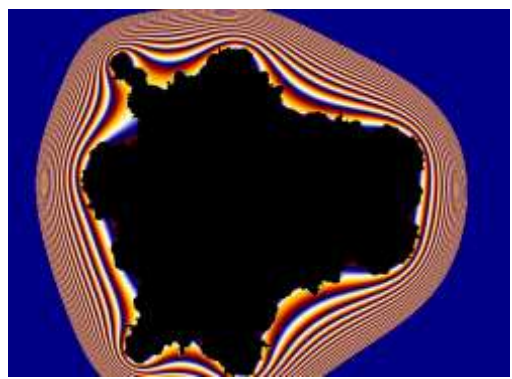

Fig 3: Relative Superior Julia Set for $\alpha =0.4$, $\beta =0.6$, c = 0.1

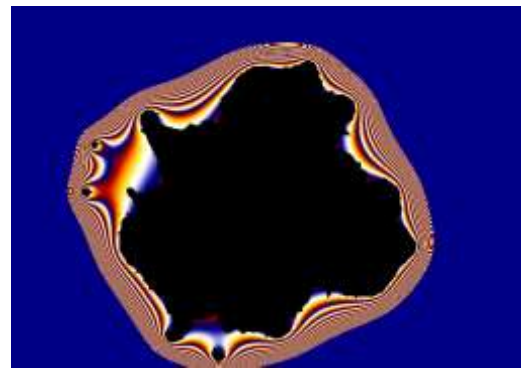

Fig 3: Relative Superior Julia Set for $\alpha = 0.4$, $\beta = 0.6$, c=0.1

**7. Generation of Relative Superior Mandelbrot Sets**

We generated the Relative Superior Mandelbrot sets. We present here some beautiful filled Relative Superior Mandelbrot sets for quadratic, cubic and biquadratic function.

1.7 Relative Superior Mandelbrot sets for Quadratic function

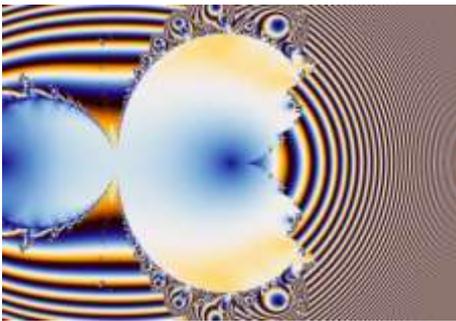

Fig1: Relative Superior Mandelbrot Set for $\alpha = \beta = 0.5$ & c = 0.1

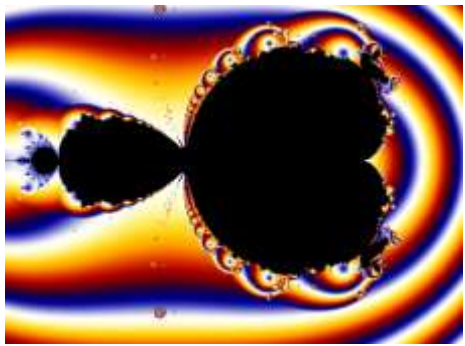

Fig 2: Relative Superior Mandelbrot Set for $\alpha = 0.8$, $\beta = 0.4$, c=0.1

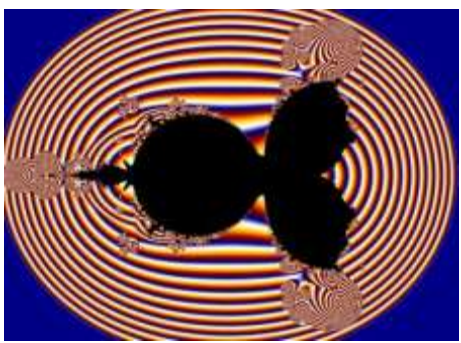

1.8 Relative Superior Mandelbrot Sets for Cubic function

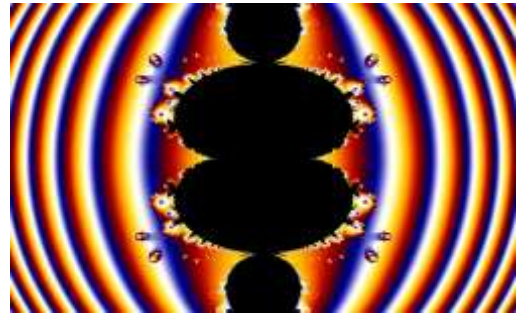

Fig 1: Relative Superior Mandelbrot Set for $\alpha = \beta = 0.8$, c=0.1

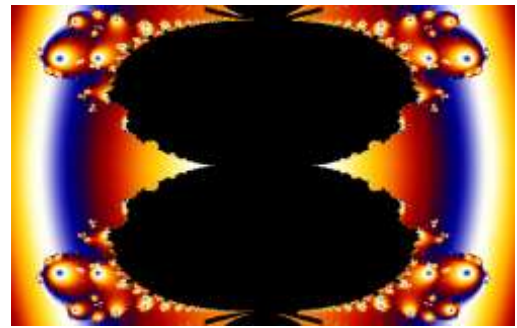

Fig 2: Relative Superior Mandelbrot Set for $\alpha = 0.5$, $\beta = 0.5$, c = 0.1

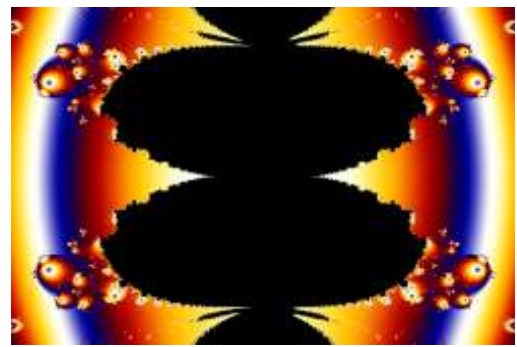

Fig 3: Relative Superior Mandelbrot Set for $\alpha = 0.4$, $\beta = 0.6$, c = 0.1

1.9 Relative Superior Mandelbrot sets for biquadratic function:

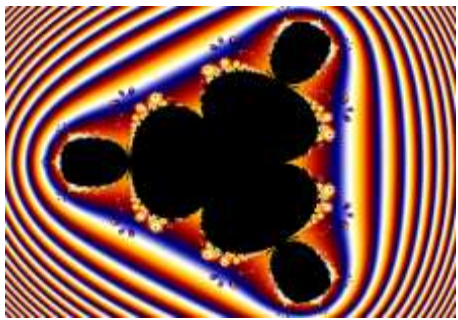

Fig 1: Relative Superior Mandelbrot Set for $\alpha$ =0.8, $\beta$ =0.8, c = 0.1

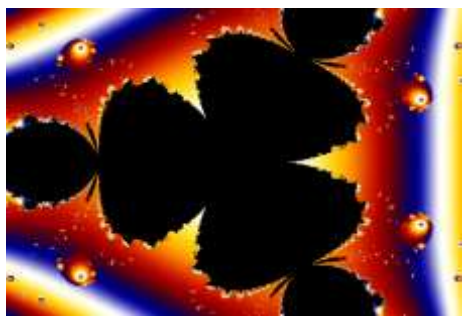

Fig 2: Relative Superior Mandelbrot Set for $\alpha$ =0.5, $\beta$ =0.5, c = 0.1

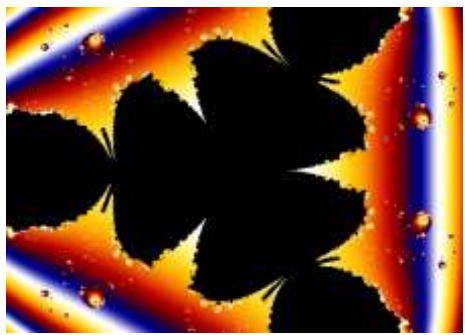

Fig 3: Relative Superior Mandelbrot Set for $\alpha$ =0.4, $\beta$ =0.6, c = 0.1

## IV. CONCLUSION

In the dynamics of complex polynomial $z^n - z + c = 0$ for $n \geq 2$, all the Relative Superior Mandelbrots are symmetrical objects, and for even values of (n) all the Relative Superior Mandelbrots are symmetrical about x-axis and for odd values of (n) all the Relative Superior Mandelbrots are symmetrical about both axis(x-axis and y-axis).